\documentclass[traditabstract]{aa}
\usepackage{txfonts}
\usepackage{amssymb}
\usepackage{graphicx}
\def\sign={\rm sign}
\begin{document}
\title{Constructing the secular architecture of the solar system I: The giant planets}
\author{A. Morbidelli \inst{1}
\and R. Brasser \inst{1}
\and K. Tsiganis \inst{2}
\and R. Gomes \inst{3}
\and H. F. Levison \inst{4}}
\institute{Dep. Cassiopee, University of Nice - Sophia Antipolis, CNRS, Observatoire de la C\^{o}te d'Azur; Nice, France
\and Department of Physics, Aristotle University of Thessaloniki; Thessaloniki, Greece
\and Observat\'{o}rio Nacional; Rio de Janeiro, RJ, Brasil
\and Southwest Research Institute; Boulder, CO, USA}
\date{submitted: 13 July 2009; accepted: 31 August 2009}
\abstract{Using numerical simulations, we show that smooth migration of the giant planets through a planetesimal disk leads to an orbital architecture that is inconsistent with the current one: the 
resulting eccentricities and inclinations of their orbits are too small. The crossing of 
mutual mean motion resonances by the planets would excite their orbital eccentricities but not their orbital inclinations. Moreover, the amplitudes of the eigenmodes characterising the current secular evolution of the eccentricities of Jupiter and Saturn would not be reproduced correctly; only one eigenmode is excited by resonance-crossing. We show that, at the very least, encounters between Saturn and one of the ice giants (Uranus or Neptune) need to have occurred, in order to reproduce the current secular properties of the giant planets, in particular the amplitude of the two strongest 
eigenmodes in the eccentricities of Jupiter and Saturn.}
\titlerunning{Secular architecture of the giant planets}
\maketitle

\section{Introduction}
The formation and evolution of the solar system is a longstanding open
problem. Of particular importance is the issue of the origin of the
orbital eccentricities of the giant planets. Even though these are
small compared to those of most extra-solar planets discovered so far,
they are nevertheless large compared to what is expected from
formation and evolution models. \\

Giant planets are expected to be born on quasi-circular orbits because
low relative velocities with respect to the planetesimals in the disk
are a necessary condition to allow the rapid formation of their cores
(Kokubo \& Ida, 1996, 1998; Goldreich {{\textit{et al.}}}, 2004).
Once the giant planets have formed, their eccentricities evolve under
the effects of their interactions with the disc of gas.  These
interactions can in principle enhance the eccentricities of very
massive planets (Goldreich \& Sari, 2003), but for moderate-mass
planets they have a damping effect. In fact, numerical hydro-dynamical
simulations (Kley \& Dirksen, 2006; D'Angelo {{\textit{et al.}}},
2006) show that only planets of masses larger than 2--3 Jupiter masses
that are initially on circular orbits are able to excite an
eccentricity in the disk and, in response, to become eccentric
themselves. Planets of Jupiter-mass or less have their eccentricities
damped. Accounting for turbulence should not change the result
significantly: the eccentricity excitation due to turbulence is only
of the order of 0.01 for a 10 Earth mass planet and decreases rapidly
with increasing mass of said planet (Nelson, 2005). By comparison, the
mean eccentricities of Jupiter and Saturn are 0.045 and 0.05
respectively. \\

\begin{figure}
\resizebox{\hsize}{!}{\includegraphics[angle=-90]{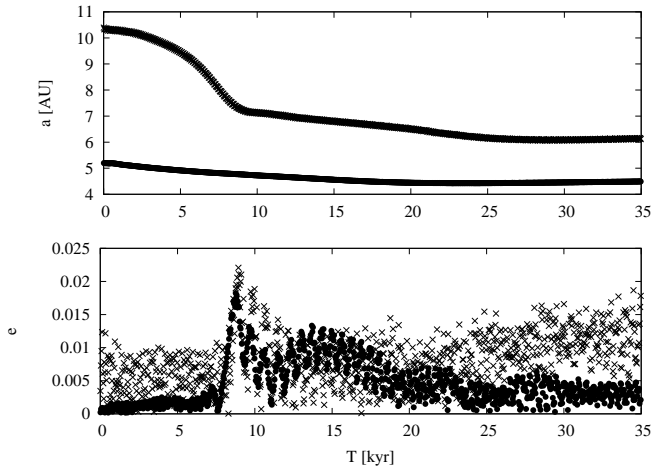}}
\caption{The evolution of Jupiter (bullets) and Saturn (crosses) in the gas disk. 
Taken from Morbidelli \& Crida (2007), but reproducing the evolution
shown in Masset \& Snellgrove (2001).}
\label{MassSnell} 
\end{figure}

In addition, the interactions between Jupiter and Saturn, as they
evolve and migrate in the disk of gas, should not lead to a
significant enhancement of their
eccentricities. Figure~\ref{MassSnell} shows a typical evolution of
the Jupiter-Saturn pair, from Masset \& Snellgrove (2001). The top
panel shows the evolution of the semi major axes, where Saturn'
semi-major axis is depicted by the upper curve and that of Jupiter is
the lower trajectory. Initially far away, Saturn swiftly approaches
Jupiter, possibly passing across their mutual 2:1 resonance (at
approximately 9\,000~yr in the figure), and is eventually trapped in
the 3:2 resonance. At this point, the migration of both planets slows
down slightly and then reverses.  Morbidelli \& Crida (2007) argued
that this dynamical evolution explains why Jupiter did not migrate all
the way to the Sun in our System. Pierens \& Nelson (2008)
convincingly demonstrated that the trapping in the 3:2 resonance is
the only possible outcome for the Jupiter-Saturn pair. The lower panel
of Figure~\ref{MassSnell} shows the evolution of the eccentricities of
both planets, where Saturn's eccentricity is depicted by crosses and
that of Jupiter by bullets.  Both eccentricities remain low all the
time. The burst of the eccentricities associated to the passage
through the 2:1 resonance at approximately 9\,000~yr is rapidly
damped. Once trapped in the 3:2 resonance, the equilibrium
eccentricities are approximately 0.003 for Jupiter and 0.01 for
Saturn, i.e. five to ten times smaller than their current values. \\

\begin{figure}
\resizebox{\hsize}{!}{\includegraphics[angle=-90]{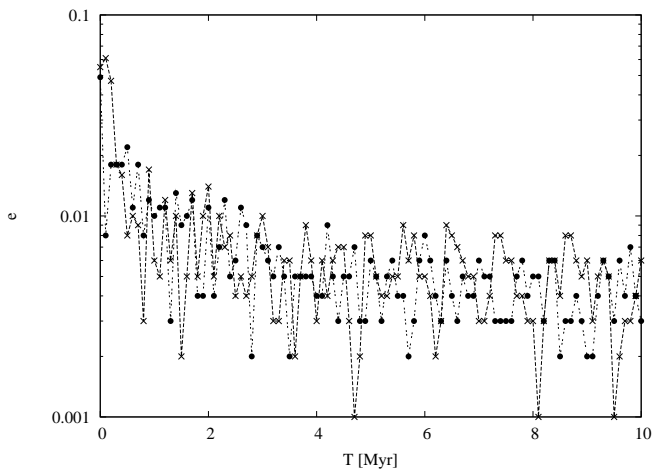}}
\caption{The evolution of the eccentricities of Jupiter (bullet) and Saturn (crosses)  
as they migrate through a 50 Earth masses planetesimal disk. 
From Gomes {{\textit{et al.}}} (2004).}
\label{renu} 
\end{figure}

Once the gas has dispersed from the system, the giant planets are
still expected to migrate, due to their interaction with a
planetesimal disk (Fernandez \& Ip, 1984; Malhotra, 1993, 1995; Hahn
\& Malhotra, 1999; Gomes {{\textit{et al.}}}, 2004). While migration
in the gas disk causes the planets to approach each other (Morbidelli
{{\textit{et al.}}}, 2007), migration in the planetesimal disk causes
the planets to diverge i.e.  it increases the ratio between the
orbital periods (Fernandez \& Ip, 1984).  In this process, the orbital
eccentricities are damped by a mechanism known as ``dynamical
friction'' (e.g. Stewart \& Wetherill, 1988). Figure~\ref{renu}
provides a example of the eccentricity evolution of Jupiter (bullets)
and Saturn (crosses) who are initially at 5.4 and 8.7 AU with their
current eccentricities (initial conditions typical of Malhotra, 1993,
1995; Hahn \& Malhotra 1999).  They migrate, together with Uranus and
Neptune, through a planetesimal disk carrying in total
50$M_{\oplus}$. The disk is simulated using 10,000 tracers (see Gomes
{{\textit{et al.}}}, 2004, for details). The figure shows the
evolution of the eccentricities of Jupiter and Saturn: both are
rapidly damped below 0.01.  Thus, a smooth radial migration through
the planetesimal disk, as originally envisioned by Malhotra (1995)
cannot explain the current eccentricities (nor the inclinations) of
the orbits of the giant planets.  \\

Then, how did Jupiter and Saturn acquire their current eccentricities?
In Tsiganis {{\textit{et al.}}} (2005), the foundation paper for a
comprehensive model of the evolution of the outer Solar System --
often called the {\it Nice} model -- it is argued that the current
eccentricities were achieved when Jupiter and Saturn passed across
their mutual 2:1 resonance, while migrating in divergent directions
under the interactions with a planetesimal disk. They indeed showed
that the {\it mean} eccentricities of Jupiter and Saturn are
adequately reproduced during the resonance crossing (see electronic
supplement of Tsiganis {{\textit{et al.}}}, 2005, or figure~\ref{c21}
below), as well as their orbital separations and mutual
inclinations. However, the mean values of the eccentricities do not
properly describe the secular dynamical architecture of a planetary
system: the eccentricities of the planets oscillate with long periods,
because of the mutual secular interactions among the planets. A system
of $N$ planets has $N$ fundamental frequencies in the secular
evolution of the eccentricities, and the amplitude of each mode -- or,
at least that of the dominant ones -- should be reproduced in a
successful model. We remind that Tsiganis {{\textit{et al}}}. (2005) never checked if the
Nice model reproduces the secular architecture of the giant planets (we
will show below that it does) nor if this is achieved via the 2:1
resonance crossing (we will show here that it is not). \\

Actually, in this paper we make an abstraction of the Nice model, and investigate
which events in the evolution of the giant planets are needed to
achieve the current secular architecture of the giant planet system.
We start in section~2 by reviewing what this secular architecture is
and how it evolves during migration, in the case where no mean motion
resonances are crossed. In section 3 we investigate the effect of the
passage through the 2:1 resonance on the secular architecture of the
Jupiter-Saturn pair. As we will see, this resonance crossing alone,
although reproducing the mean eccentricities of both planets, does not
reproduce the frequency decomposition of the secular system. In section~4 we
discuss the effect of multiple mean motion resonance crossings between
Jupiter and Saturn, showing that this is still not enough to 
achieve the good secular solution. In section~5 we examine the role of a third
planet, with a mass comparable to that of Uranus or Neptune. We
first consider the migration of this third planet on a circular orbit,
then on an eccentric orbit and finally we discuss the consequences of
encounters between this planet and Saturn. We show that encounters of 
Saturn with the ice giant lead to the correct secular evolution for 
the eccentricities of Jupiter and Saturn. In section~6 we return
to the Nice model, verify its ability to reproduce the current
secular architecture of the planetary system and discuss other
models that could in principle be equally successful in this
respect. Although this paper is mostly focused on the Jupiter-Saturn 
pair and the evolution of their eccentricities, in section~7 we briefly
discuss the fate of Uranus and Neptune and the excitation of 
inclinations. The case of the terrestrial planets will be discussed in a second
paper. The results are then summarised in section~8.

\section{Secular eccentricity evolution of the Jupiter-Saturn pair}
\label{secsys}

One can study the secular dynamics of a pair of planets as
described in Michtchenko \& Malhotra (2004). In that case the
planets are assumed to evolve on the same plane and are far from
mutual mean motion resonances. The Hamiltonian describing their
interaction is averaged over the mean longitudes of the planets. This
averaged Hamiltonian describes a two-degrees-of-freedom system, whose
angles are the longitudes of perihelia of the two planets: $\varpi_1$
and $\varpi_2$. The D'Alembert rules (see Chapter 1 of
Morbidelli, 2002), ensure that the Hamiltonian depends only on the combination
$\Delta\varpi\equiv\varpi_1-\varpi_2$. Thus the system is
effectively reduced to one degree of freedom, which is
integrable. This means that, in addition to the value of the averaged
Hamiltonian itself, which will improperly be called ``energy''
hereafter, the system must have a second constant of motion. Simple
algebra on canonical transformations of variables 
allows one to prove that this constant is
\begin{equation}
K=m_1'\sqrt{\mu_1 a_1} \left(1-\sqrt{1-e_1^2}\right) + 
m_2'\sqrt{\mu_2 a_2} \left(1-\sqrt{1-e_2^2}\right)\ ,
\label{Kint}
\end{equation}
where $\mu_i=G(M+m_i)$, $m_i'=m_i M/(M+m_i)$, $M$ is the mass of the
star, $G$ is the gravitational constant and $m_i, a_i$ and $e_i$ are
the mass, semi major axis an eccentricity of each planet. Note that
$K$, called {\it angular momentum deficit}, actually measures the
deviation in angular momentum for an eccentric two-planet system,
with respect to a system of two planets, both on circular orbits, with
the same values of $a_i$. Now the global secular dynamics of the
system can be illustrated by plotting level curves of the energy over
manifolds defined by the condition that $K$ is constant.

\begin{figure}
\resizebox{\hsize}{!}{\includegraphics[angle=-90]{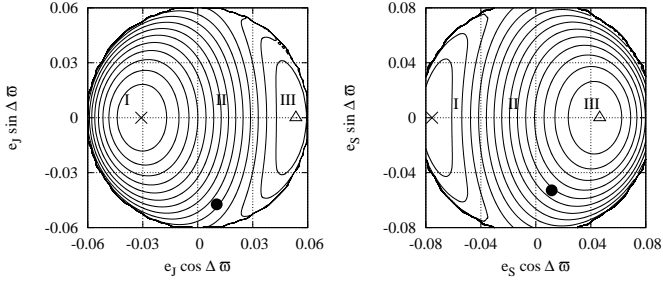}}
\caption{Global illustration of the secular dynamics of the
  Jupiter-Saturn system. The bullets
  represent the current values of $e_J$, $e_S$ and $\Delta\varpi$. 
}
\label{mmphasejsnow} 
\end{figure}

Figure~\ref{mmphasejsnow} shows the result for the Jupiter-Saturn
system. The value of $K$ that we have chosen corresponds to the
current masses, semi major axes and eccentricities of these planets.
The left panel illustrates the dynamics in the coordinates
$e_J\cos\Delta\varpi$ and $e_J\sin\Delta\varpi$, while the right panel
uses the coordinates $e_S\cos\Delta\varpi$ and $e_S\sin\Delta\varpi$,
where $e_J$ refers to the eccentricity of Jupiter and $e_S$ to the
eccentricity of Saturn.  The bullets represent the current
configuration of the Jupiter-Saturn system. We stress that the two
panels are just two representations of the {\it same} dynamics. The
same level curves of the energy are plotted in both panels. Thus, the
$n^{{{\rm{th}}}}$ level curve counting from the triangle in the left
panel corresponds to the $n^{{{\rm{th}}}}$ level curve counting from
the triangle in the right panel. Indeed, the dot representing the
current Jupiter-Saturn configuration is on the 5$^{{{\rm{th}}}}$ level
curve away from the triangle on each panel. The secular evolution of
the system has to follow the energy level curve that passes through
the dot. The other energy curves show the secular evolution that
Jupiter and Saturn would have had, if the system were modified
relative to the current configuration, preserving the current value of
$K$. We warn the reader that the dynamics illustrated in this figure is not
very accurate from a quantitative point of view, because we have
neglected the effects of the nearby 5/2 mean motion resonance between
Jupiter and Saturn.  Nevertheless all the qualitative aspects of the
real dynamics are correctly reproduced. \\

We remark that the global secular dynamics of the Jupiter-Saturn
system is characterised by the presence of two stable equilibrium
points, one at $\Delta\varpi=0$ (marked by a triangle in
figure~\ref{mmphasejsnow}) and one at $\Delta\varpi=\pi$ (marked by a
cross). Thus, there are three kinds
of energy level curves along which the Jupiter-Saturn system could
evolve: those along which $\Delta\varpi$ librates around $\pi$ (type
I), those along which $\Delta\varpi$ circulates (i.e. assumes all
values from 0 to $2\pi$; type II) and those along which $\Delta\varpi$
librates around $0$ (type III). Notice that
while type II curves wrap around the stable equilibrium at
$\Delta\varpi=\pi$ in the left panel, the curves wrap around the stable
equilibrium at $\Delta\varpi=0$ in the right panel. This means that
during the circulation of $\Delta\varpi$, the eccentricity of Jupiter
has a maximum when $\Delta\varpi=\pi$ while that of Saturn has a
maximum when $\Delta\varpi=0$. The real Jupiter-Saturn system has this
type of evolution. \\

We stress that there is no critical curve (separatrix) separating the
evolutions of type I, II and III. By critical curve we mean a
trajectory passing though (at least) one unstable equilibrium point,
along which the travel time is infinite; an example is the curve
separating the libration and circulation regimes in a pendulum. In
this respect, speaking of ``resonance'' when $\Delta\varpi$ librates,
as it is sometimes done when discussing the secular dynamics of
extra-solar planets, is misleading because the word ``resonance'', in
the classical dynamical systems and celestial mechanics terminology,
implies the existence of such a critical curve.

\begin{table}
\begin{tabular}{ccc}
 Frequency & Value ($\arcsec$/yr) & Phase ($\degr$)\\ 
\\\hline\\
$g_5$ &  4.26 &  30.67 \\
$g_6$ & 28.22 & 128.11 \\
 \\\hline \\
\end{tabular}
\caption{Frequencies and phases for the secular evolution of Jupiter
  and Saturn on their current orbits.
}
\label{sjs}
\end{table}

\begin{table}
\begin{tabular}{c|cc}
$_j\setminus^k$\quad\hbox{} & 5& 6 \\
\\\hline\\
5 & 0.0442& 0.0157\\ 
6 &  0.0330& 0.0482\\  
 \\\hline \\
\end{tabular}
\caption{Coefficients $M_{j,k}$ of the Lagrange--Laplace solution 
for the Jupiter-Saturn system. The coefficients of the terms with
frequencies other than $g_5$ and $g_6$ are omitted.
}
\label{sjs2}
\end{table}

In addition to using phase portraits, the secular dynamics of the
Jupiter-Saturn system, or any pair of planets with small
eccentricities, can also be described using the classical
Lagrange-Laplace theory (see Chapter 7 in Murray \& Dermott, 1999).
This theory, which is in fact the solution of the averaged problem
described above, in the linear approximation, states that the
eccentricities and longitudes of perihelia of the pair of planets
evolve as:
\begin{eqnarray}
e_J\cos\varpi_J&=&M_{5,5}\cos\alpha_5-M_{5,6}\cos\alpha_6\cr 
e_J\sin\varpi_J&=&M_{5,5}\sin\alpha_5-M_{5,6}\sin\alpha_6\cr 
e_S\cos\varpi_S&=&M_{6,5}\cos\alpha_5+M_{6,6}\cos\alpha_6\cr 
e_S\sin\varpi_S&=&M_{6,5}\sin\alpha_5+M_{6,6}\sin\alpha_6
\label{Lagrange}
\end{eqnarray}
where $\alpha_5=g_5 t + \beta_5$ and $\alpha_6=g_6 t + \beta_6$. Here
$g_5$ and $g_6$ are the eigenfrequencies of the system, while
$\beta_5$ and $\beta_6$ are their phases at $t=0$. In equation
(\ref{Lagrange}) all $M_{j,k} >0$.  Tables~\ref{sjs} and~\ref{sjs2}
report the values of all the coefficients, obtained from the Fourier
analysis of the complete 8-planet numerical solution (Nobili {\it
{{\textit{et al.}}}}, 1989).  \\

There is a one-to-one correspondence between the relative amplitudes
of the coefficients $M_{j,k}$ and the three types of secular evolution 
illustrated in figure~\ref{mmphasejsnow}. We detail this relationship
below, in order to achieve a better understanding of the planetary
evolutions illustrated in the next sections. \\

From equation (\ref{Lagrange}) the evolution of $e_J\cos\Delta\varpi$ and 
$e_S\cos\Delta\varpi$ (the quantities plotted on the $x$-axes of the
panels in figure~\ref{mmphasejsnow}) are:
\begin{eqnarray}
 e_J\cos\Delta\varpi&=&\left[M_{5,5}M_{6,5}-M_{5,6}M_{6,6}\right.\cr
&&\left.+(M_{5,5}M_{6,6}-M_{5,6}M_{6,5})\cos(\alpha_5-\alpha_6)\right]/e_S\cr
 e_S\cos\Delta\varpi&=&\left[M_{5,5}M_{6,5}-M_{5,6}M_{6,6}\right.\cr
&&\left.+(M_{5,5}M_{6,6}-M_{5,6}M_{6,5})\cos(\alpha_5-\alpha_6)\right]/e_J
\label{ecosdelta}
\end{eqnarray}
where
\begin{equation}
e_J=\sqrt{M_{5,5}^2+M_{5,6}^2-2M_{5,5}M_{5,6}\cos(\alpha_5-\alpha_6)}
\label{eJ}
\end{equation}
and
\begin{equation}
e_S=\sqrt{M_{6,5}^2+M_{6,6}^2+2M_{6,5}M_{6,6}\cos(\alpha_5-\alpha_6)}\ .
\label{eS}
\end{equation}
When $\alpha_5-\alpha_6=0$ one has
\begin{eqnarray}
 e_J\cos\Delta\varpi&=&(M_{5,5}-M_{5,6}){\rm sign} (M_{6,5}+M_{6,6})\cr
 e_S\cos\Delta\varpi&=&(M_{6,5}+M_{6,6}){\rm sign} (M_{5,5}-M_{5,6})\ ,
\label{ecos0}
\end{eqnarray}
where sign() is equal to $-1$ if the argument of the function is
negative, $+1$ if it is positive and 0 if it is zero. Instead, when
$\alpha_5-\alpha_6=\pi$ one has
\begin{eqnarray}
 e_J\cos\Delta\varpi&=&(M_{5,5}+M_{5,6}){\rm sign} (M_{6,5}-M_{6,6})\cr
 e_S\cos\Delta\varpi&=&(M_{6,5}-M_{6,6}){\rm sign} (M_{5,5}+M_{5,6})\ .
\label{ecosPi}
\end{eqnarray}
Now, suppose that the amplitude corresponding to the $g_5$ frequency is zero,
i.e.\ $M_{5,5}=M_{6,5}=0$. Then, the dependence of equation (\ref{ecosdelta}) on
$\alpha_5-\alpha_6$ vanishes and, from equation (\ref{ecos0}) or
equation (\ref{ecosPi}), one sees that the system is located at the
equilibrium point at $\Delta\varpi=\pi$, so that $e_J=M_{5,6}$ and
$e_S=M_{6,6}$ (the point marked by a cross in figure~\ref{mmphasejsnow}). \\

Let us now gradually increase the
amplitudes of the $g_5$ mode, relative to that of the $g_6$ one. This implies increasing $M_{5,5}$ and $M_{6,5}$ at the same rate, while keeping $M_{5,6}$ and $M_{6,6}$ fixed.
Initially, when $M_{5,5}$ and $M_{6,5}$ are small compared to $M_{5,6}$ and 
$M_{6,6}$, all quantities
in equations (\ref{ecos0}) and (\ref{ecosPi}) are negative, and therefore the
evolution of the system follows an energy level curve of type I, along
which $\Delta\varpi$ librates around $\pi$. The distance of this curve
from the equilibrium point, which we call {\it amplitude of oscillation} hereafter, 
is directly proportional to $M_{5,5}$ or $M_{6,5}$.  \\

Since $M_{5,6} < M_{6,6}$  and $M_{6,5} < M_{5,5}$, then when $M_{5,5} = M_{5,6}$ one has $M_{6,5} < M_{6,6}$. This implies that increasing the amplitude of the $g_5$ mode eventually brings us to the situation where $M_{5,5}$ becomes larger than $M_{5,6}$, but $M_{6,5}$ is still less than $M_{6,6}$. Now the value of
$e_J\cos\Delta\varpi$ at $\alpha_5-\alpha_6=0$
i.e.\ $M_{5,5}-M_{5,6}$ becomes positive. When additionally $\alpha_5-\alpha_6=\pi$ 
its value remains negative, i.e. $-(M_{5,5}+M_{5,6})$. Thus, the system now evolves on 
an energy curve of type II,
along which $\Delta\varpi$ circulates. Notice that the value of
$e_S\cos\Delta\varpi$ at $\alpha_5-\alpha_6=\pi$ remains negative, 
while the value at $\alpha_5-\alpha_6=0$ jumps from 
$-(M_{6,5}+M_{6,6})$ to $(M_{6,5}+M_{6,6})$. Thus, the level curve in
the right panel of figure~\ref{mmphasejsnow} flips from one looping
around the equilibrium point at $\Delta\varpi=\pi$ to one looping around the
equilibrium at $\Delta\varpi=0$.  \\

Further increasing the amplitude of the $g_5$ mode
relative to that of the $g_6$ mode, eventually results in $M_{6,5}$ also becoming
larger than $M_{6,6}$. Now, all quantities in equations (\ref{ecos0}) and
(\ref{ecosPi}) are positive, which means the the system follows an
energy level curve of type III, along which $\Delta\varpi$ librates
around 0. Notice that the value of $e_J\cos\Delta\varpi$ at
$\alpha_5-\alpha_6=\pi$ jumps from $-(M_{5,5}+M_{5,6})$ to
$(M_{5,5}+M_{5,6})$, which means that the the level curve in the left
panel of figure~\ref{mmphasejsnow} flips from one going around the
equilibrium point at $\Delta\varpi=\pi$ to one going around the equilibrium
at $\Delta\varpi=0$. \\

Finally, when the amplitude of the $g_6$ mode is zero, the
system is on the stable equilibrium at $\Delta\varpi=0$ (the cross in
figure~\ref{mmphasejsnow}).  \\

Below we discuss how the secular dynamics of the planets changes as
they migrate away from each other and are { also} submitted to 
dynamical friction, exerted by the planetesimal population.

\subsection{Migration and the evolution of the secular dynamics}
\label{migr}

Let's imagine two planets migrating, without
passing through any major mean motion resonance. A good example could
be Jupiter and Saturn migrating from a configuration with orbital
period ratio $P_S/P_J$ slightly larger than 2 to their current
configuration, with $P_S/P_J$ slightly smaller than 2.5. \\

The migration causes the semi-major axis ratio between the planets 
to change. This affects the values of the coefficients 
$M_{j,k}$, since they depend explicitly on the above ratio; in turn, 
this affects the global portrait of secular dynamics. However, if the migration 
is slow enough, the amplitude of oscillation around the equilibrium 
point is preserved as an {\it adiabatic invariant} (Neishtadt, 1984; Henrard, 
1993). More precisely, it can be demonstrated that the conserved quantity is 
\begin{equation}
J=\oint  m'_J\sqrt{G\mu_Ja_J} \left(1-\sqrt{1-e_J^2}\right){\rm
  d}\Delta\varpi\ ,
\label{adiabat}
\end{equation}
which is the action conjugate to $\Delta\varpi$, and
where $\oint$ denotes the integral over a closed energy curve (i.e.\ a bounded trajectory) that
characterises the secular motion of the two planets (as in
figure~\ref{mmphasejsnow}), if migration is frozen. Therefore during migration
the planets would react to the slow changes in the global dynamical
portrait, by passing from one energy curve to another in such a way as to
preserve the quantity in equation (\ref{adiabat}), i.e.\ the oscillation
amplitude around the stable equilibrium point remains constant. Since for 
Jupiter and Saturn this amplitude is related to $M_{5,5}$, it turns out that 
any smooth migration should not have changed this coefficient significantly. 
In the next section the additional effect of dynamical friction is analysed.

\subsection{Dynamical friction and the evolution of the secular dynamics}
\label{damp}

Dynamical friction is the mechanism by which gravitating objects of
different masses exchange energy so as to evolve towards an
equipartition of energy of relative motion (Saslaw, 1985). For a
system of planets embedded in a massive population of small bodies, 
the eccentricities and inclinations of the former are damped, while those of 
the latter are excited (Stewart \& Wetherill, 1988). \\

In principle, each planet might suffer dynamical friction with 
different intensity, owing to a different location inside the
planetesimal disk. However, the planets are connected to each other
through their secular dynamics, so that dynamical friction, even if
acting in an unbalanced way between the planets, turns out to have a
systematic net effect. \\

\begin{figure}
\resizebox{\hsize}{!}{\includegraphics[angle=-90]{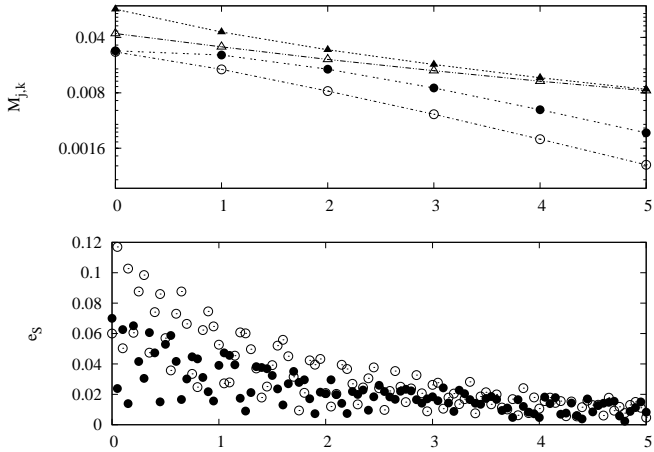}}
\caption{Effect of eccentricity damping on the evolution of Jupiter
  and Saturn. In this experiment the damping force is applied to
  Saturn only. The top panel shows the coefficients $M_{5,5}$ (filled circles), $M_{5,6}$ (open circles), $M_{6,5}$
(filled triangles) and $M_{6,6}$ (open triangles) as a function of
  time. The bottom panel shows the eccentricities of Saturn (open
  circles) and Jupiter (filled circles).}
\label{fig-damp} 
\end{figure}

To illustrate this point, consider again the Jupiter-Saturn system of
figure~\ref{mmphasejsnow} and suppose that dynamical friction is applied
only to Saturn. The eccentricity of Saturn is damped, so the value of
$K$ is reduced. Consequently, the location of the two stable
equilibrium points has to move towards $e=0$. If the adiabatic
invariance of equation (\ref{adiabat}) held, the amplitude of oscillation
around the equilibrium point { would} be preserved, eventually turning a
libration of $\Delta\varpi$ into a circulation. However, the adiabatic
invariance does not hold in this case. The reason is that dynamical
friction damps the eccentricity of Saturn, and therefore damps both
the $M_{6,6}$ {\it and} the $M_{6,5}$ coefficients. Since the $M_{5,5}$ and  
$M_{6,5}$ coefficients are related, $M_{5,5}$ is also damped. In other
words, the amplitude of oscillation around the equilibrium point is
damped, and so the value of $J$ given in equation (\ref{adiabat}) decays with time. \\

As a check, we have run a simple numerical experiment. We have
considered a Jupiter-Saturn system with semi major axes 5.4 and 8.85,
with relatively eccentric orbits and large amplitude ($60\degr$) of
apsidal libration around $\Delta\varpi=180\degr$. We have integrated
the orbits using the Wisdom-Holman (Wisdom \& Holman, 1991) method, with the code
Swift-WHM (Levison \& Duncan,1994). We used a time-step of 0.1~y and
modified the equations of motion so that a damping term is included
for the eccentricity of Saturn only. Figure~\ref{fig-damp} shows the
result. The bottom panel shows the evolutions of the eccentricities
of the two planets, where $e_S$ is represented by open circles and
$e_J$ by filled circles: both are damped and decay with time at the
same rate. The top panel shows the amplitudes of the coefficients of
equation (\ref{Lagrange}), i.e.\ $M_{5,5}$ (filled circles), $M_{5,6}$
(open circles), $M_{6,5}$ (filled triangles) and $M_{6,6}$ (open
triangles), computed at six different points in time. The
computations were performed, by applying Fourier analysis to the time
series produced in respectively six short-time integrations, where no
eccentricity damping was applied. As one can see, all coefficients
decrease with time, at comparable rates.\\

Thus, we conclude that, whatever the initial configuration of the
planets, smooth migration and dynamical friction cannot increase the
amplitude of the $g_5$ term and cannot turn the libration of
$\Delta\varpi$ into a circulation. This result will be relevant in the
next section. 

\section{The effect of the passage through the 2:1 resonance}
\label{s21}

We now consider the migration of Jupiter and Saturn, initially on
quasi-circular orbits, through their mutual 2:1 mean motion
resonance. Tsiganis {{\textit{et al.}}} (2005) argued that this passage through the
resonance is responsible for the acquisition of the current
eccentricities of the two planets. \\

\begin{figure}
\resizebox{\hsize}{!}{\includegraphics[angle=-90]{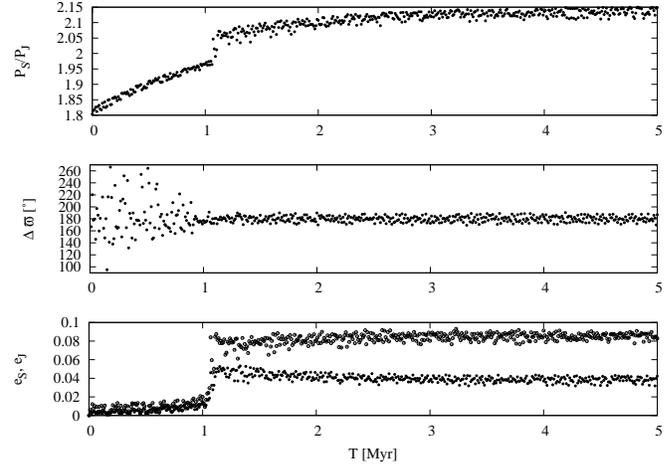}}
\caption{The evolution of Jupiter and Saturn, as they pass across
  their mutual 2:1 resonance. In the bottom panel, Saturn's
  eccentricity is the upper curve and Jupiter's is the lower one.}
\label{c21} 
\end{figure}

Figure \ref{c21} shows the effect of the passage through this
resonance, starting from circular orbits. The simulation is  again done
using the Swift-WHM integrator, but in this case 
the equations of motion are modified so as to induce radial migration
to the planets, with a rate decaying as $\exp(-t/\tau)$. No
eccentricity damping is imposed. In practise, at { every}
timestep $h$ the velocity of each planet is multiplied by a quantity
$(1+\beta)$, with $\beta$ being proportional to $h\exp(-t/\tau)$. For
Jupiter $\beta$ is negative and for
Saturn it is positive, so that the two planets migrate inwards and outwards 
respectively, as observed in realistic $N$-body simulations (Fernandez 
\& Ip, 1984; Hahn \& Malhotra,
1999; Gomes {{\textit{et al.}}}, 2004). We choose $\tau=1$ Myr. \\

\begin{figure}
\resizebox{\hsize}{!}{\includegraphics[angle=-90]{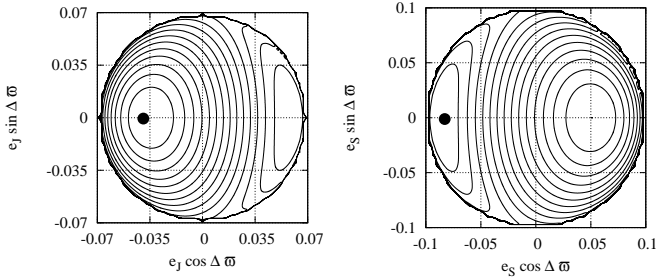}}
\caption{Similar to figure~\ref{mmphasejsnow} but after the 2:1 resonance crossing.}
\label{mmphasejs21} 
\end{figure}

The top panel of Figure \ref{c21} shows the ratio of the orbital
periods of Saturn ($P_S$) and Jupiter ($P_J$) as a function of
time. We stop the simulation well before $P_S/P_J$ achieves the
current value, to emphasise the effect of the 2:1 resonance
crossing. The middle panel shows $\Delta \varpi$ as a function of
time. The bottom panel shows the evolutions of the eccentricities of
Jupiter and Saturn, where the lower trajectory corresponds to Jupiter
and the upper curve corresponds to Saturn. We notice that the orbital
period ratio abruptly jumps across the value of 2. Correspondingly,
the eccentricities of Saturn and Jupiter jump to $\sim 0.07$ and $\sim
0.045$, which, as noticed by Tsiganis {{\textit{et al.}}} (2005), are
quite close to the current mean eccentricities of the two
planets. During the subsequent migration, the eccentricity of Saturn
increases somewhat and that of Jupiter decreases respectively. The
two planets enter into apsidal anti-alignment (i.e.\ $\Delta\varpi$
librates around 180 degrees) shortly before the resonance passage and
the libration amplitude shrinks down to $\sim 10^\circ$ as the
eccentricities o the two planets grow. The crossing of the 2:1 does
not seem to significantly affect the libration amplitude, which
remains of the order of $\sim 10^\circ$ during the
post-resonance-crossing migration (see also \v{C}uk, 2007). As a result of this narrow
libration amplitude, the eccentricities of Jupiter and Saturn do not
show any sign of secular oscillation. In practise, Jupiter and Saturn
are located at the stable equilibrium point of their secular dynamics,
as shown in figure~\ref{mmphasejs21}. This is very different from the
current situation (compare with
figure~\ref{mmphasejsnow}). Table~\ref{sjs21} reports the values of
the $M_{i,k}$ coefficients of equation (\ref{Lagrange}) at the end of
the simulation. The $M_{5,5}$ and $M_{6,5}$ coefficients, related to
the amplitude of oscillation around the equilibrium point as explained
in section~\ref{secsys}, are very small; they are more than an order
of magnitude smaller than their current values. As discussed in
section~\ref{migr}, they would not increase during the
subsequent migration of the planets, because they behave as adiabatic
invariants. Even worse, they would decrease if dynamical friction
were applied. \\

\begin{table}
\begin{tabular}{c|cc}
$_j\setminus^k$\quad\hbox{} & 5& 6 \\
\\\hline\\
5 & 0.00272& 0.00275\\ 
6 &  0.0378& 0.0854\\  
 \\\hline \\
\end{tabular}
\caption{Coefficients $M_{j,k}$ of the Jupiter-Saturn secular system at the
  end of the simulation illustrated in figure~\ref{c21}.}
\label{sjs21}
\end{table}

\begin{table}
\begin{tabular}{c|c}
$\tau$ [Myr] & $M_{5,5} \times 10^{-3}$ \\
\\\hline\\
1 & 3.55\\ 
2 &  4.41\\
5 & 7.49 \\
10 & 7.02 \\
20 & 5.05 
 \\\hline \\
\end{tabular}
\caption{Values of $M_{5,5}$ in Jupiter after the 2:1 resonance crossing with Saturn as a function of the migration e-folding time, $\tau$.}
\label{M55}
\end{table}

Simulations that we performed assuming larger values of $\tau$
(i.e. slower migration rates) lead to the same result. The
eccentricities of Jupiter and Saturn after the 2:1 resonance crossing
are about the same as in figure~\ref{c21}. The amplitude of libration of
$\Delta\varpi$ is always between 6 and 15 degrees, with no apparent
correlation on $\tau$. Table~\ref{M55} recapitulates the results, for
what concerns the values of the $M_{5,5}$ coefficient, which are
always comparably small. Thus, we conclude that the 2:1 resonance
crossing, although it explains the current {\it mean} values of the
eccentricities of Jupiter and Saturn, cannot by itself explain the 
current secular dynamical structure of the system. \\

Let us now provide an interpretation of the behaviour observed in
the simulation. The dynamics of two planets in the vicinity of a mean motion 
resonance can be studied following Michtchenko {{\textit{et al.}}} (2008). 
For the 2:1 resonance, the fundamental angles of the problem are
\begin{eqnarray}
\sigma_J&=&\lambda_J-2\lambda_S+\varpi_J\cr
\Delta\varpi&=&\varpi_J-\varpi_S\ .
\label{angles12}
\end{eqnarray}
The motion of the first angle is conjugated with the motion of the
angular momentum deficit $K$ of the planets, defined in equation
(\ref{Kint}). The motion of the second angle is conjugated with the
motion of the quantity $Q_S=m'_S\sqrt{G\mu_Sa_S}
\left(1-\sqrt{1-e_S^2}\right)$. If the system is far from the
resonance, the motion of $\sigma_J$ can be averaged out. Then $K$
becomes a constant of motion and the secular dynamics described in
the previous section is recovered. In particular, if the planets migrate slowly enough so that the
adiabatic invariance of $J$ -- see equation (\ref{adiabat}) -- holds, the amplitude of oscillation around the stable
equilibrium point of the secular dynamics has to remain constant. However, as migration continues,
the approach to the mean motion resonance forces the location of the equilibrium point to shift to larger eccentricity. Thus, by virtue of a geometric effect, the apparent amplitude of libration of $\Delta\varpi$ has to shrink as the eccentricities
increase. This is visible in Fig.~\ref{c21} in the phase before the resonance crossing. \\

As the planets are approaching the resonance, the angle $\sigma_J$ can
no longer be averaged out in a trivial way. However, an adiabatic
invariant can still be introduced, as long as the timescale
for the motion of $\sigma_J$ is significantly shorter than that of
$\Delta\varpi$ and that migration changes the system on even longer
time scales. This invariant is
\begin{equation}
{\cal K}=\oint K{\rm d}\sigma_J\ ,
\end{equation}
where the integral is taken over a path describing the coupled evolution
of $K$ and $\sigma_J$, which is closed { if} $Q_S$ and
$\Delta\varpi$ are frozen (i.e.\ a trajectory of the so-called ``frozen'' 
system).  \\

As shown in Michtchenko {{\textit{et al.}}} (2008), for a pair of planets with
the Jupiter-Saturn mass ratio and $e_J<0.08$, the dynamics of the 2:1
resonance presents one critical curve, or separatrix, for the
$K,\sigma_J$ degree of freedom and no critical curve for the $Q_S,
\Delta\varpi$ degree of freedom. Thus, when the resonance is reached
during the migration, the invariance of ${\cal K}$ is broken
(Neishtadt, 1984), since the two time-scales of the motion are no more 
well separated. As the planets are migrating in divergent
directions, they cannot be trapped in the resonance (Henrard \&
Lemaitre, 1983). The resonant angle $\sigma_J$ has to switch from
clockwise to anti-clockwise circulation. Correspondingly, the
quantity ${\cal K}$ has to jump to a larger value, that is the planets
acquire an angular momentum deficit. How this new angular
momentum deficit is partitioned between the two planets is difficult
to compute a priori, because the dynamics are fully four-dimensional 
(i.e. two degrees of freedom) and, therefore, not integrable. \\

\begin{figure}
\resizebox{\hsize}{!}{\includegraphics{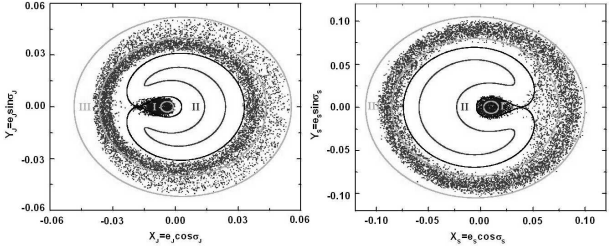}}
\caption{Left: the evolution of $\sigma_J$ and $e_J$ in polar
  coordinates. Right: the evolution of $\sigma_S$ and $e_S$. Before the 
resonance is reached, both planets have nearly zero eccentricities (region I, black 
dots). When the resonance is reached, both planets jump to the corresponding Region 
III, along the $x$-axis and through the ``X''-point of the critical
  curve (grey dots). From  the supplementary material of Tsiganis {{\textit{et al.}}}
(2005)}
\label{21portraits} 
\end{figure}

As shown in figure~\ref{21portraits}, there are clearly two regimes of
motion in the portraits $Q_J, \sigma_J$ and $Q_S,
\sigma_S\equiv\lambda_J-2\lambda_S+\varpi_S$ (remember that $e_J^2\sim
Q_J$ and $e_S^2\sim Q_S$). Before the resonance crossing (black dots),
the dynamics are confined in a narrow, off-centred region close to the
origin of the polar coordinates. After the resonance crossing (grey
dots), the dynamics fill a wide annulus, also asymmetric relative to
the axis $\cos\sigma=0$.  The curves in each panel are free hand
illustrations of the dynamics near and inside a first-order mean
motion resonance. The region filled by the black dots is 
bounded by the inner loop of the critical curve (labelled I in the
plot). The annulus filled by the grey dots is adhesive, at its inner
edge, to the outer loop of the critical curve. The jump in
eccentricity observed at the resonance crossing corresponds to
the passage from the region inside the inner loop to that outside the
outer loop.  

Thus, in practise, it is as if each planet saw its own resonance:
the one with critical angle $\sigma_J$ for Jupiter and the one with
critical angle $\sigma_S$ for Saturn. The two resonances are just two
slices of the same resonance, because only one critical curve exists
(Michtchenko {{\textit{et al.}}}, 2008). This is the reason why the
eccentricities of both planets jump simultaneously. The 2:1 resonance
is structured by the presence of a periodic orbit, along which
$\sigma_J$ and $\Delta\varpi$ remain constant and are equal to 0 and
$\pi$ respectively (Michtchenko {{\textit{et al.}}}, 2008). As
$\Delta\varpi=\sigma_J-\sigma_S$, { the} phase portrait of the
$\sigma_J$ resonance and that of the $\sigma_S$ resonance are rotated
by 180 degrees, with respect to each other. Thus, $Q_J$ reaches a
maximum when $\sigma_J=0$ and $Q_S$ when $\sigma_S=\pi$. Consequently,
when the planets reach their maximal eccentricities and the $\sigma$
angles start to circulate anti-clockwise, $\Delta\varpi$ has to be
$\sim 180^\circ$ (see figure~\ref{21portraits}).  It is evident that
the result of this transition through the resonance depends just on
the resonance topology and not on the migration rate, as long as the
latter is slow compared to the motion of the $\sigma$ angles (i.e.\
the adiabatic approximation holds; Neishtadt, 1984). \\

After the resonance crossing, one can { again} average over
$\sigma_J$ and reduce the system to a one-degree of freedom
secular system. As $\Delta\varpi=\pi$, Jupiter has to be on the
negative $x$-axis of a diagram like that of the left panel of
figure~\ref{mmphasejs21}. Thus the secular evolution of Jupiter will be
an oscillation around the stable equilibrium at $\Delta\varpi=\pi$.
The amplitude of this oscillation depends on the value of the
eccentricity of Jupiter acquired at resonance crossing, relative
to the value of the stable equilibrium of the secular { problem}. 
It turns out that, for the masses of Jupiter
and Saturn these two values are almost the same. Thus, the amplitude
of oscillation is very small. We think that this is a coincidence and
that, in principle, it does not have to be
that way. In fact, we have verified numerically that the result depends
on the individual masses of both planets, even for the 
same mass ratio. For instance, if the masses of Jupiter and Saturn are 
both reduced by a factor 100, the eccentricities of both planets jump to 
$\sim 0.01$ at resonance crossing, and this puts Jupiter 
on a secular trajectory that brings $e_J$ down to 0; that is, the secular 
motion is now at the boundary between type-I and type-II, as defined in
section~\ref{secsys}. This is caused by a different scaling of the jumps in
$e_J$ and $e_S$ with respect to the planetary masses and to a 
different global secular dynamics in the vicinity of the resonance, which in turn is
caused by a different relative importance of the quadratic terms in the
masses. \\

Once the planets are placed relative to the portrait of their secular
dynamics, their destiny is fixed. As they move away from the mean
motion resonance, the secular portrait can change, in particular
because the near-resonant perturbation terms that are quadratic 
in the masses rapidly decrease in amplitude. Hence the location of the 
equilibrium points can change, but the planets have to follow them 
adiabatically. This explains the slow monotonic growth of the eccentricity 
of Saturn and the decay of that of Jupiter, observed in the top panel 
of figure~\ref{c21}, while the amplitude of libration of $\Delta\varpi$ does 
not change (middle panel).

\section{Passage through multiple resonances}
\label{multiple}

Since the passage of Jupiter and Saturn through the 2:1 resonance,
starting from initially circular orbits, produces a secular system
that is incompatible with the current one, we now explore the effects
of the passage of these planets through a {\it series} of resonances.
This is done to determine whether or not such evolution could
increase the value of $M_{j,5}$ in both planets.  \\

We set Jupiter and Saturn initially on quasi-circular orbits just outside
their mutual 3:2 resonance. The choice of these initial conditions is
motivated by the result that during the gas-disk phase, Saturn should
have been trapped in the 3:2 resonance with Jupiter (Morbidelli {{\textit{et
al.}}}, 2007; Pierens \& Nelson, 2008). Once the gas disappeared
from the system, the two planets should have been extracted from the
resonance at low eccentricity, by the interaction with the
planetesimals, and subsequently start to migrate. \\

\begin{figure}
\resizebox{\hsize}{!}{\includegraphics[angle=-90]{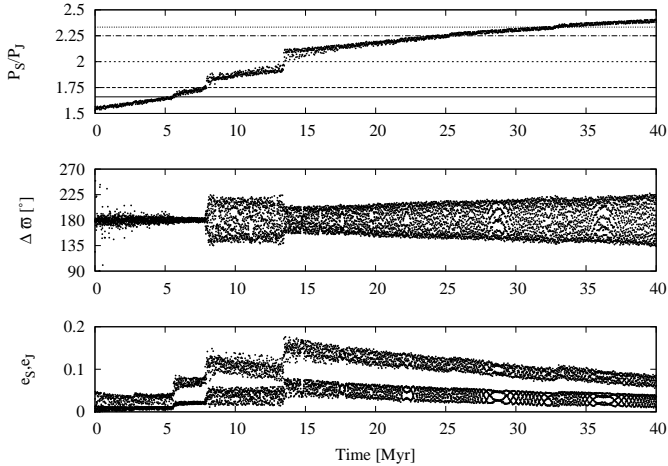}}
\caption{Like figure~\ref{c21}, but for Jupiter and Saturn evolving
  through a sequence of mean motion resonances, from just outside
  their mutual 3:2 commensurability up to their current location.}
\label{c532173} 
\end{figure}

In the above setting, our planets are forced to migrate through the
5:3, 7:4, 2:1, 9:4 and 7:3 resonances, ending up close to their
current location in semi major axis (i.e.\ slightly interior to the
5/2 resonance).  In the migration equations we set $\tau$ so that
it takes about 40~Myr to reach $P_S/P_J\approx 2.5$ although, as we saw
before, the migration timescale has little influence on the resonant
effects. The result of this experiment is shown in
figure~\ref{c532173}, which is similar in format to
figure~\ref{c21}.  The horizontal lines in the top panel denote the
positions of the resonances mentioned above. Notice a distinct jump
in the eccentricities of both planets at each resonance crossing. In
order to prevent the system from becoming unstable, we applied
eccentricity damping to Saturn, so to mimic the effect of dynamical
friction and reach final eccentricities that are similar to the
current mean values of the two planets. The parameters for the simulation 
depicted in figure~\ref{c532173} are $\tau= 25$ Myr and $\dot{e}_S = -2 \times 10^{8}$ yr$^{-1}$. 
The effect of damping is visible in the eccentricity evolution, after the 2:1 resonance
crossing; we will discuss the effect on the motion of
$\Delta\varpi$ below.\\

The middle panel of figure~\ref{c532173} shows that the passage through
the 7:4 resonance significantly increases the libration amplitude of
$\Delta\varpi$. In fact, the amplitude of the $M_{5,5}$ term in
(\ref{Lagrange}), increases from $9\times 10^{-4}$ before the
resonance crossing, to $0.019$ after the crossing. The
passage through the 2:1 resonance, however, shrinks the amplitude of
libration of $\Delta\varpi$. This happens because the 2:1 resonance crossing, as
we have seen in the previous section, does not enhance $M_{5,5}$ (it
remains equal to $0.019$ in this simulation) but
does enhance the overall eccentricities of the planets. As explained earlier, this causes the amplitude 
of libration of $\Delta\varpi$ to decrease. \\

Notice from figure~\ref{c532173} that, after the 2:1 resonance
crossing, the amplitude of libration of $\Delta\varpi$ starts to
increase, slowly and monotonically. This is caused { not} by an enhancement
of the amplitude of oscillation around the equilibrium point of the
secular dynamics, but by the damping of the eccentricities of the
planets. It is the opposite of what was just described before: a
geometrical effect. In reality, the value of $M_{5,5}$ is {\it
decreased} to $0.015$ (from 0.019) during this evolution. Hence, at
the end of the simulation, the amplitude of the $g_5$ mode is 
about a factor of 3 smaller than in the real secular dynamics of Jupiter 
and Saturn. Without eccentricity damping, the amplitude of the $g_5$
mode would have remained equal to $\sim 0.019$, still much smaller
than in the current Jupiter-Saturn secular dynamics.\\

Several other experiments, changing the initial conditions slightly or
the migration speed $\tau$, lead essentially to the same result. Thus,
we conclude that the migration of Jupiter and Saturn through a 
sequence of mean motion resonances is not enough to achieve their
current secular configuration. A richer dynamics is required,
likely involving interactions with a third planet. 

\section{Three-planet dynamics}

From the discussions and the examples reported above, it is quite
clear that, to enhance the amplitudes of the $g_5$ mode, it is
necessary that the eccentricity of Saturn receives a kick that is not
counterbalanced by a corresponding increase in the eccentricity of
Jupiter (or vice versa). This would indeed move the planets away from
the stable equilibrium point of their secular dynamics, thus enhancing
the amplitude of oscillation around this point and, consequently,
$M_{j,5}$. Given that Jupiter and Saturn are not alone in the outer
solar system, in this section we investigate the effect that 
interactions with {{\textit{a third planet with a mass comparable to that of Uranus and Neptune, which we simply refer to as 'Uranus'}}}, has on the Jupiter-Saturn pair. 
We first address the effects of the migration of Uranus on a 
quasi-circular orbit. Then we study the effects of its migration on an 
initially eccentric orbit and, finally, we address the problem of 
encounters among the planets.

\subsection{Migration of Uranus on a quasi-circular orbit}

The main mean motion resonance with Saturn that Uranus can go through
is the 2:1. Thus, this is the resonance crossing that we focus on
here.  Given that, as we have seen in the previous sections, the
effect of a passage through a mean motion resonance is quite
insensitive to the migration rate, the initial location of the planets
etc., the main issue that may potentially lead to different results is
whether the crossing of the 2:1 Saturn-Uranus resonance happened
before or after the putative crossing of the 2:1 Jupiter-Saturn
resonance. Below we investigate each of these two cases. \\

To have the crossing of the Saturn-Uranus resonance happen first, we have
performed a numerical experiment, with Saturn and Jupiter having
initially a small orbital period ratio ($P_S/P_J=1.53$), and Uranus 
and Saturn having an orbital period ratio $P_U/P_S\sim 1.95$.
The exact initial locations are not important, as long as they do not
change the order of the resonance crossings. 
All planets initially have circular orbits. The three
planets are forced to migrate to their current positions with $\tau=5$
Myr. Eccentricity damping is imposed on Saturn and Uranus, to mimic
dynamical friction, with forces tuned such that, at the end of the
simulation, Uranus approximately reaches its current eccentricity. \\

\begin{figure}
\resizebox{\hsize}{!}{\includegraphics[angle=-90]{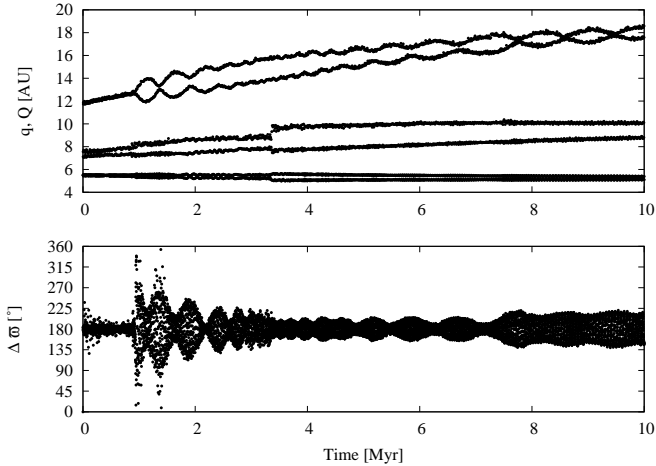}}
\caption{A three-planet migration simulation, in which both the 2:1 resonance
  between Uranus and Saturn and the 2:1 resonance between Saturn and
  Jupiter are crossed.}
\label{US21} 
\end{figure}

Figure~\ref{US21} shows the result. The top panel shows the
pericentre $q$ and apocentre $Q$ of the planets which, from top to
bottom, are Uranus, Saturn and Jupiter respectively. The separation
among these curves gives a visual measure of the eccentricity of the
orbit of the respective planet. The middle panel
shows $\Delta \varpi$ for Jupiter and Saturn. \\

In this simulation, Uranus crosses the 2:1 resonance with Saturn at
$t\sim 0.9$ Myr. This gives a kick to the eccentricity of Uranus (its
$q,Q$-curves abruptly separate from each other) and, to a lesser
extent, to the eccentricity of Saturn. This sudden increase in the
eccentricity of Saturn moves the stable equilibrium point of the
Jupiter-Saturn secular dynamics away from $e_J\sim e_S\sim
0$. However, the eccentricity of Jupiter does not receive an
equivalent kick by this resonance crossing, so it remains close to
zero. Consequently, Jupiter must start evolving secularly along a
trajectory close to the boundary between type-I and type-II curves
(see section~\ref{secsys}); in other words $M_{5,5}\sim M_{5,6}$. This
is the reason why the amplitude of libration of $\Delta\varpi$
changes abruptly at the Uranus-Saturn resonance crossing, and reaches
an amplitude of $\sim 180^\circ$. \\

The interim between 0.9 and 3.3~Myr is characterised by large, long-periodic,
oscillations of the eccentricity of Uranus, which correlate with the
modulation of the amplitude of libration of $\Delta\varpi$. These
oscillations have a frequency equal to $g_5-g_7$, where $g_7$ is
the new fundamental frequency that characterises the extension of
(\ref{Lagrange}) to a three-planet system. Soon after the
Uranus-Saturn resonance crossing, $g_5-g_7$ is
small and therefore the oscillations have large amplitude. As Uranus
departs from the resonance with Saturn, $g_7$ decreases; at the
same time, $g_5$ increases, since Jupiter approaches the 2:1 
resonance with Saturn. Hence, the oscillation with frequency $g_5-g_7$ 
becomes faster and its amplitude deceases. This sequence of increasingly 
shorter oscillations reduces the overall amplitude of libration of 
$\Delta\varpi$ to approximately 40 degrees. \\

At $t\sim 3.3$~Myr, Jupiter and Saturn cross their mutual 2:1 mean
motion resonance, which has the effects that we discussed before. The
amplitude of $M_{5,5}$ is roughly preserved in this resonance
crossing, as already illustrated in section~\ref{multiple}. Its value at
$t=4$~Myr is 0.010, much larger than in section~\ref{s21} but still about
a factor of four smaller than the current value. The $M_{5,5}$ 
coefficient receives an additional small enhancement at $t\sim 7.5$~Myr, 
when Jupiter and Saturn cross their 7:3 resonance, but this does not change 
the substance of the result. \\

To reverse the order of the resonance crossings, we have run a second
experiment, in which we placed Jupiter and Saturn just beyond their
2:1 resonance ($P_S/P_J = 2.06$), on orbits typical of those achieved
by the 2:1 resonance crossing (see section~\ref{s21}). This means
that Jupiter and Saturn are in apsidal libration around 180$\degr$,
with an amplitude of the $g_5$ mode that is small in both planets,
compared to the current value (see table~\ref{sjs21}). Uranus was
placed on a circular orbit at $a_U=12.5$. Again the planets were
forced to migrate to their current locations, with $\tau=5$ Myr. The
2:1 resonance crossing between Uranus and Saturn again kicked the
eccentricity of Saturn, which in turn enhanced the amplitude of the
$g_5$ term. In this case, the final value of the $M_{5,5}$ coefficient
was 0.014, i.e.\ of the same order as in the previous
experiment. \\

Given the above results, we conclude that, no matter when the
Uranus-Saturn 2:1 resonance crossing occurs, there is an enhancement
of the amplitude of the $g_5$ term, as expected, but it is too small (by a factor of
$\sim 3$) to explain the current Jupiter-Saturn secular system. It appears that the mass of Uranus is too small to provide enough
eccentricity excitation on Saturn, when passing through a mean
motion resonance with it.

\subsection{Migration of Uranus on an eccentric orbit}
 
In the current solar system, the proper frequency of perihelion of
Uranus, $g_7$, is smaller than $g_5$ (3.1 and 4.3 $\arcsec$/yr,
respectively). If Uranus was much closer to Saturn, however, $g_7$ had to be 
much larger too. For instance, if Uranus were just outside the 2:1 resonance with Saturn (say
$a_U=14.8$~AU and $a_S=8.6$~AU), $g_7$ was $\sim 6.5\arcsec$/yr.
One might then wonder if, during the migration of Uranus the $g_5=g_7$ secular resonance 
could have occurred. On the other hand, if Jupiter was closer to Saturn, $g_5$ would 
have been higher as well; $g_5> 6.5\arcsec$/yr if $P_S/P_J\lesssim
2.2$. Thus, the occurrence of the $g_5=g_7$ resonance depends on
the locations of both Jupiter and Uranus, relative to Saturn. \\

\begin{figure}
\resizebox{\hsize}{!}{\includegraphics[angle=-90]{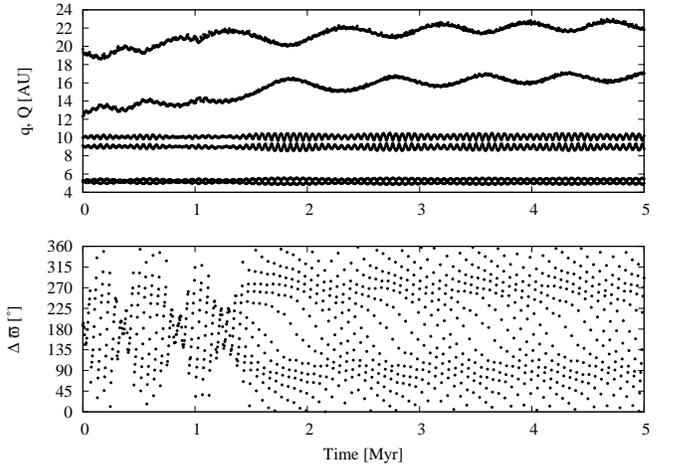}}
\caption{A three-planet simulation, with Uranus initially on an
  eccentric orbit and migrating to its current location. No migration
  is imposed on Jupiter and Saturn.}
\label{eccu} 
\end{figure}

To see the effect of this secular resonance, we performed an idealised
experiment, in which we placed Jupiter at $a_J=5.2$~AU, Saturn at
$a_S=9.5$~AU and Uranus at $a_U=16$~AU with an eccentricity of 0.25.
The initial eccentricities and $\Delta\varpi$ for Jupiter and Saturn
were taken from a run, in which these planets passed through their 
mutual 2:1 resonance and migrated up to the locations reported above.
Hence $\Delta\varpi$ would librate with very small amplitude,
in the absence of Uranus. In this experiment the latter was forced to migrate
towards its current location, with $\tau=2$ Myr. No migration was
imposed on Jupiter and Saturn. Since initially $g_5=4.4\arcsec$/yr
and $g_7>g_5$, the $g_5=g_7$ resonance crossing had to occur during
this simulation. No eccentricity damping was applied to any of the
planets in this run. \\

The result is shown in figure~\ref{eccu}. In the first part of the
simulation ($t<1.3$~Myr) the amplitude of libration of $\Delta\varpi$,
which is initially very small, suffers a large modulation,
correlated with the oscillations of the eccentricity of Uranus. The
dynamics here are in analogous to what we described before, for the
interim between the two mean motion resonance crossings in
figure~\ref{US21}. At $t\sim 1.3$~Myr, the $g_5=g_7$ resonance is
crossed. This leads to an exchange of angular momentum between Uranus
and Jupiter. The eccentricity of Uranus decreases a bit, while the
value of the $M_{5,5}$ coefficient is enhanced. As a response,
$\Delta\varpi$ starts to circulate. The final value of $M_{5,5}$ is
0.04, essentially matching the current value. \\

Although Jupiter has to be far from Saturn to have a genuine secular
resonance crossing, we have found that more realistic simulations,
in which Jupiter and Saturn are initially much closer to each other
-- so to be able to migrate in the correct proportion with respect to 
Uranus -- can lead to interesting results as well. The reason is that, although from the
beginning $g_7<g_5$, the two frequencies can become quasi-resonant; such 
interactions also allow for a significant transfer of
eccentricity from Uranus to Jupiter and can excite the value
of $M_{5,5}$ up to the current figure. \\

However, the reader should be aware that, while the effect of mean
motion resonances is quite insensitive on parameters and initial
conditions, in the case of a secular resonance, the outcome depends
critically on a variety of issues. More precisely, the effects of the
$g_5=g_7$ resonance, or quasi-resonance, must depend on the
eccentricity of Uranus, the migration timescale $\tau$ and the
position of $\varpi_U$ relative to $\varpi_J$, immediately before
this resonant interaction. The reason for the first dependence is
that $e_U$ sets the strength of the secular resonance. The
dependence on $\tau$ and $\varpi_U$ has to do with the fact that the
timescale associated with a secular resonance is very long, of the
order of 1~Myr. Thus, migration through a secular resonance, unlike
migration through a mean motion resonance, is not an adiabatic
process, at least for values of $\tau$ up to 10~Myr that we are
focusing on here. Thus, the time spent in the vicinity of the
resonance and the values of the phases at which the planets enter
the resonance have important impact on the resulting dynamics. \\

Given the above, we conclude that the secular interaction with an
eccentric Uranus is a mechanism that is potentially capable of
exciting the $g_5$ mode in the Jupiter-Saturn system to the observed
level, but this mechanism is quite un-generic. Moreover, if we invoked
an eccentric Uranus to explain the origin of the Jupiter-Saturn
dynamical architecture, we would still need to explain how Uranus got so
eccentric in first place. Finally, the $g_5=g_7$ secular
resonance cannot alone explain the excitation of the planetary inclinations, 
which will be discussed in section~7. For all these reasons, we continue our 
search for a better mechanism and consider below the effect of planetary encounters.

\subsection{Planetary encounters with Uranus}

Close encounters between Uranus and Saturn could potentially be a 
very effective mechanism for kicking the eccentricity of Saturn 
and enhancing the amplitude of the $g_5$ mode. \\

To investigate this, we have run a series of twenty simulations, where the
initial semi-major axes of Jupiter and Saturn were chosen such that
these two planets are just outside their 2:1 mean motion resonance
($P_S/P_J$ = 2.06), on orbits typical of those achieved during the 2:1
resonance crossing (in apsidal anti-alignment with negligible
oscillation amplitude). Uranus was placed on an orbit with semi-major axis ranging 
from 11.8~AU to 13.4~AU at 0.2~AU intervals, with an initial 
eccentricity of 0.1. This value of the eccentricity is of the order of 
that achieved by Uranus, under the secular forcing induced by
Jupiter and Saturn. The system was then allowed to evolve under the 
mutual gravitational forces and external migration forces. For all simulations, 
the e-folding time for the migration forces was set at 5 Myr. Eccentricity 
damping was applied to Uranus and Saturn, the values being 
$\dot{e}_S = -2 \times 10^{-8}$/yr and $\dot{e}_U = -1.2 \times 10^{-7}$/yr. 
The damping coefficient for Uranus was 
assumed to be six times larger than that of Saturn because, in principle,
Uranus is more affected by the planetesimal disk than the gas-giants.
The strength of the damping term was calibrated so that the post-encounter 
evolution of the eccentricity of Uranus follows the one seen in the full 
$N$-body simulations of Tsiganis {{\textit{et al.}}} (2005). These details are 
not very important, because we focus here on the final secular dynamics of
Jupiter and not on the final orbit of Uranus. The latter is very
sensitive to the prescription of damping, but not the former as we
have seen in sect.~2.2. Uranus was typically found to be scattered by Saturn (and sometimes by Jupiter). 
The simulations were stopped once the phase of encounters among
the planets ended, either because Uranus was decoupled from the giant
planets, due to eccentricity damping, or because it was ejected 
from the system. \\

\begin{figure}
\resizebox{\hsize}{!}{\includegraphics[angle=-90]{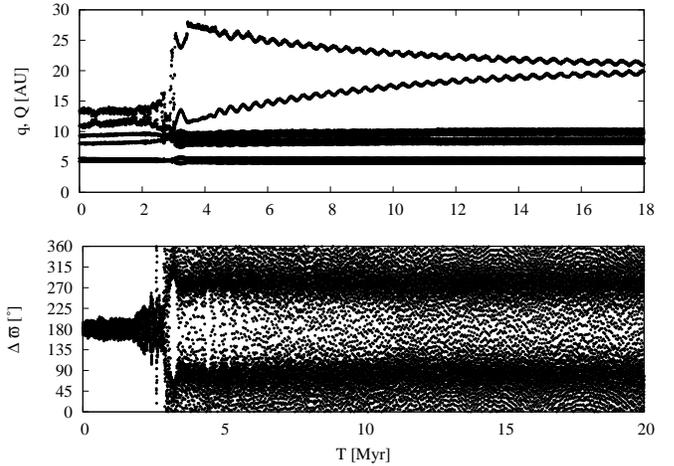}}
\caption{Example of an encounter between Uranus and Saturn. The plot is similar to figure~\ref{US21}.}
\label{encounters} 
\end{figure}

The simulations that yielded the best results in terms of 
final $M_{5,5}$ value are those with initial semi-major
axes of Uranus $a_U < 13$ AU. In these successful simulations, a total of four or 20\%, 
the average final value of $M_{5,5}$ was approximately
0.04, in very good agreement with the current configuration. 
Figure~\ref{encounters} gives an example of evolution from one of 
these successful runs, and is similar to figure~\ref{US21} and 
figure~\ref{eccu}. As seen in our results, initially placing Uranus 
further away from Saturn results in encounters that are too weak to pump 
up the $g_5$ mode in Jupiter. \\

In summary, we conclude that encounters between Saturn and Uranus
constitute an effective and quite generic mechanism for achieving a
final secular evolution of the Jupiter-Saturn system that is consistent
with their current state. Compared to all other mechanisms
investigated in this paper, which either do not work or work only for
an ad-hoc set of conditions, planet-planet scattering is our
favoured solution to the problem of the origin of the secular
architecture of the giant planet system. In Section~\ref{incl} 
we provide further arguments in favour of this conclusion. 

\section{The Nice model and its alternatives}

The work presented above shows that a combination of the effects
provided by the 2:1 resonance crossing between Jupiter and Saturn and 
by encounters and/or secular interactions with an eccentric Uranus, can
produce a Jupiter-Saturn system that behaves secularly like the real
planets. \\

The 2:1 resonance crossing, the encounters among the planets and the
high-eccentricity phases of Uranus and Neptune are essential
ingredients of the Nice model (Tsiganis {{\textit{et al.}}}, 2005;
Gomes {{\textit{et al.}}}, 2005; Morbidelli {{\textit{et al.}}},
2007). Thus, we expect that this model not only reproduces the mean
orbital eccentricities of the planets, as shown in Tsiganis
{{\textit{et al.}}} (2005), but also the correct architecture of secular
modes. Curiously, this has never been properly checked before, and
we do so in the following. \\

\begin{figure}
\resizebox{\hsize}{!}{\includegraphics[angle=-90]{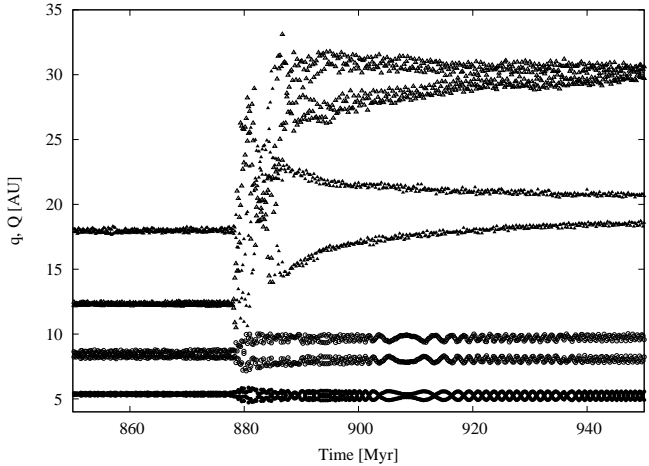}}
\caption{Sample evolution of the giant planets in the
  {{\textit{Nice}}} model (from Gomes {{\textit{et al.}}}, 2005). Each planet is
  represented by a pair of curves, showing the time evolution of their
  perihelion and aphelion distance.}
\label{nice1} 
\end{figure}
 
In Figure \ref{nice1}, the pericentre and apocentre distance of the
four giant planets are plotted as a function of time, in a
simulation taken from Gomes {{\textit{et al.}}} (2005) that
adequately reproduces the current positions of the giant planets. The
curve starting around 5 AU represents Jupiter, the one around 9 AU is
Saturn, the trajectory at 12 AU is Uranus (who ends up switching
positions with Neptune) and the uppermost curve at 16 AU is Neptune
(who ends up closer to the Sun than Uranus). This plot is a
magnification around the time when Jupiter and Saturn cross their 2:1
resonance and the system becomes unstable. The plot shows the phase
until all encounters had stopped, which in this case happened when the
period ratio between Jupiter and Saturn was $P_S/P_J = 2.23$. The
final semi-major axes of the four planets are
$(a_J,a_S,a_U,a_N)=(5.23, 8.94, 19.88, 31.00)$, so that the largest
``error'' is in Saturn's orbit. A Fourier spectrum of Jupiter's
eccentricity at the end of the simulation gives $M_{5,5}=0.027$, and
$M_{5,6}=0.036$. The amplitude of the $g_5$ term is a bit small, but
well within a factor of 2 from the real value. Another Nice-model run
from the same Gomes {{\textit{et al.}}} (2005) study gave an
essentially identical result. However, a third simulation gave
$M_{5,5}=0.059$, which is higher than the real value. In a fourth
simulation, in which not only Saturn but also Jupiter encountered an
ice giant, a value close to the real one was again recoverd, namely
$M_{5,5}=0.037$.  Given the chaotic nature of planetary encounters
and that the resulting $M_{5,5}$ values are close to the real one
(0.044) or even larger (e.g. 0.059), we conclude that the Nice model
is able to reproduce the secular architecture of the Jupiter-Saturn
system. \\

In principle, depending on the initial separations between Saturn and
the innermost ice giant, the Nice model can also give planetary
evolutions in which encounters between Saturn and an ice giant do not
occur (only the ice giants encountering each other; see Tsiganis {{\textit{et
al.}}}, 2005). These evolutions have been rejected already in Tsiganis
{{\textit{et al}}}. (2005), because they lead to final mean eccentricities (and
inclinations) that are too small and an orbital separation between
Saturn and Uranus that is too narrow et the end. They also lead to a
value of $M_{5,5}$ that is much too small compared to the current
value, because the 2:1 resonance crossing alone is not capable of
pumping the excitation of the $g_5$ mode, as we have seen earlier. \\

At this point, one might wonder whether the Nice model, in the version
with Saturn-Uranus encounters, is the only
model capable of this result. From the study reported in this paper,
it seems likely that encounters among planets might be sufficient to
excite the modes of the final secular system, without any need for 
Jupiter and Saturn crossing their mutual 2:1 resonance. In
other words, one might envision a model where Jupiter and Saturn
formed on circular orbits, well separated from each other in the beginning, 
so that $P_S/P_J$ was always larger than 2. These planets then had close 
encounters with other planetary embryos, which at the end left them on 
eccentric orbits with both the $g_5$ and $g_6$ modes excited. \\

A single encounter of an embryo with one planet would not work
because, by kicking the eccentricity of one planet and not of the
other, it would produce a secular system with $M_{5,5}\sim M_{5,6}$
(if the embryo encountered Saturn) or with $M_{6,6}\sim M_{6,5}$ (if
the embryo encountered Jupiter). The real system is different from
these two extremes. However, multiple encounters with one planet or
with both of them should do the job. To achieve an estimate of the
mass of the planetary embryo that could excite the secular modes of
Jupiter and Saturn up to the observed values, we have run four sets of
four simulations each. In each run we considered Jupiter, Saturn and
one embryo, initially on circular orbits. The mass of the embryo was
1, 5, 10 and 15 Earth masses respectively, for the four sets of
simulations. The initial location of the embryo was $a_e=7.2$~AU,
8.0~AU, 10.1~AU, 10.7~AU, for the four simulations in each set,
whereas Jupiter and Saturn were initially at $a_J=5.4$ AU and
$a_S=8.9$ AU in all cases. In most simulations, the embryo was
eventually ejected from the Solar System: in two runs the embryo
collided with Jupiter. The values of the $M_{5,5}$ and $M_{6,6}$
coefficients for each set of simulations are reported in
table~\ref{Thommes}. It turns out that the putative embryo had to be
massive, of the order of $>10~M_{\oplus}$. We stress that multiple embryos with the same total mass would not do an equal job, because the geometries of the encounters would be randomized, rather leading to dynamical friction instead of excitation. In fact, we did the same experiment with 100 Mars mass objects instead of a unique 10 Earth-mass object; it resulted in $M_{5,5}$ and $M_{6,6}$ being smaller than 0.001, demonstrating that an ensemble of small objects could not have excited the relevant modes to their current states. In conclusion, for the excitation of the $M_{5,5}$ mode to reach its current value, Jupiter and Saturn
should have encountered Uranus or Neptune or a putative third ice giant of comparable mass. Therefore, for what concerns the excitation of the
secular Fourier modes of the planetary orbits, a generic scenario of
global instability and mutual scattering of the four giant planets, as
originally proposed by Thommes {{\textit{et al}}}. (1999) would work;
the passage of Jupiter and Saturn through their mutual 2:1 resonance,
which is specific to the Nice model relative to Thommes {{\textit{et
al}}}. (1999) (or Thommes {{\textit{et al}}}., 2007, in which Jupiter
and Saturn are initially locked in the 2:1 resonance) is not
necessary. \\

Pierens and Nelson (2008), however, showed that the only possible
final configuration achieved by Jupiter and Saturn in the gas disk is  
in their mutual 3:2 resonance. Unless alternative evolutions have been
missed in that work, this result invalidates the initial planetary
configurations considered in Thommes {{\textit{et al}}}. (1999; 2007) and supports 
the Nice model, in particular in its newest
version described in Morbidelli {{\textit{et al.}}} (2007), where
the four giant planets start locked in a quadruple resonance with
$P_S/P_J=3:2$. 

\begin{table}
\begin{tabular}{cc|cc}
$m_e$ [$M_{\oplus}$]& $a_e$ [AU] & $M_{5,5} \times 10^{-3}$ & $M_{6,6} \times 10^{-3}$\\
\\\hline\\
1 & 7.2 & 4.41 & 5.92\\ 
1 & 8 & 1.34 & 3.17 \\
1 & 10.1 & 0.475 & 8.91\\
1 & 10.7 & 8.97 & 8.56\\
5 & 7.2 & 5.23 & 85.3\\
5 & 8.0 & 15.0 & 57.3 \\
5 & 10.1 & 7.78 & 12.0\\
5 & 10.7 & 7.49 & 20.2 \\
10 & 7.2 & 38.0 & 88.5\\
10 & 8.0 & 22.5 & 15.1\\
10 & 10.1 & 11.0 & 32.8\\
10 & 10.7 & 66.0 & 28.7\\
15 & 7.2 & 32.3 & 41.0\\
15 & 8.0 & 20.3 & 28.9\\
15 & 10.1 & 20.4 & 80.9\\
15 & 10.7 & 14.8 & 152.1
 \\\hline \\
\end{tabular}
\caption{Values of $M_{5,5}$ in Jupiter and $M_{6,6}$ in Saturn after ejecting planetary
  embryos of various masses from various original locations.}
\label{Thommes}
\end{table}

\section{Ice giants and inclination constraints}
\label{incl}

Up to now we have focused our discussion on the secular evolution of the 
eccentricities of Jupiter and Saturn. However, the
outer Solar System has two additional planets: Uranus and Neptune,
which introduce the additional frequencies $g_7$ and $g_8$ in the
secular evolution of the eccentricities. Thus, one should also 
be concerned about the correct excitation of the $g_7$ and $g_8$ modes in 
all planets. In addition, the planets have a rich secular
dynamics in inclination, associated with the precession of their
nodes. The excitation of the correct modes in 
the inclinations is a problem as crucial as that of the eccentricities.\\

The reason that we did not discuss these issues so far is because
considerations based solely on the secular evolution of the
eccentricities of Jupiter and Saturn have proved enough to guide us
towards a solution, which is also valid in the more general problem. That
is, these planetary encounters that are
necessary to explain the Jupiter-Saturn secular architecture, can also 
explain the excitation of the $g_7$ and $g_8$ modes i.e.\ $M_{j,7}$ and 
$M_{j,8}$, and of the inclinations of the planets. \\

The excitation of the $g_7$ and $g_8$ modes does not appear to be very
consraining. During the encounters between Saturn-Uranus and
Uranus-Neptune encounters, the eccentricities of the ice giants become
typically much larger than the current values. Thus, the combination of
encounters and dynamical friction can produce a wide range of
amplitudes of the $g_7$ and $g_8$ modes, including the current
amplitudes. But we cannot exclude that other mechanisms, such as a
sequence of mutual resonance crossing, could have led to the correct
amplitudes, as well.\\

Conversely, the inclination excitation is particularly interesting, because it
provides a strong, additional argument in favour of a violent evolution of the
planets that involves mutual close encounters. In fact, the planets
should form on essentially co-planar orbits, for the same reasons for
which they should form on circular orbits: low relative velocities
with respect to the planetesimals in the disk are necessary for the 
rapid formation of their cores. Once the planets are formed, tidal interactions 
with the gas disk damp the residual inclinations of the planets (Lubow 
\& Ogilvie, 2001). After the disappearance of the gas, dynamical friction 
exerted by the remnant planetesimal disk would also damp the planetary
inclinations. Thus, similar to the eccentricities, a relatively-late 
excitation mechanism is required to explain the inclinations of the planets. 
However, unlike the eccentricities, the passage across mean motion resonances 
does not significantly excite the inclinations because the resonant
terms depending on the longitude of the nodes are at least quadratic
in the inclinations. Secular resonances are also 
ineffective, if all planetary inclinations are initially small. The only 
mechanism { by} which inclinations can be efficiently increased is 
by close encounters. In fact, in the Nice model, the inclinations of Saturn,
Uranus and Neptune, relative to the orbit of Jupiter, are well reproduced. A 
similar result holds also for the Thommes {{\textit{et al.}}} (1999)
model. \\

It is interesting to note that the eccentricities of the giant
planets are about twice as large as the inclinations (respectively
$\sim 0.05$ and $\sim 0.025$ radians or $\sim 1.5^\circ$). This is
what one would expect, if both the eccentricities and the inclinations 
had been acquired by a combination of gravitational scattering and dynamical 
friction. Conversely, if encounters among the planets had never happened 
and the eccentricities had been acquired through specific resonance crossings, 
we would expect the planetary eccentricities to be much larger than the inclinations.

\section{Conclusion}

In this work we have demonstrated (see Fig. 2 and sects. 2.1 and 2.2)
that the secular architecture of the giant planets, which have non
negligible eccentricities and inclinations and specific amplitudes of
the modes in the eccentricity evolution of Jupiter and Saturn, could
not have been achieved if the planets migrated smoothly through a
planetesimal disk, as originally envisioned in Malhotra (1993; see
also Malhotra 1995; Hahn \& Malhotra, 1999; Gomes {{\textit{et al.}}},
2004). Thus, we believe that the community should no longer
consider the smooth migration model as a valid template for the
evolution of the solar system. Even repeated passages through
mutual resonances, that would have happened if the planetary system
was originally more compact than envisioned in Malhotra's model, could
not account for the orbital architecture of the giant planets, as we
observe them today. \\

Instead, the outer planetary system had to evolve in a more violent
way, in which the gas giants encountered the ice giants, or other
rogue planets of equivalent mass. In this respect, the correct
templates for the planetary evolution are the Nice model or the
Thommes {{\textit{et al.}}} model (or variants of these
two). As we noted above, the Nice model seems { to be}, so
far, more coherent and consistent in all its facets.
We remark that  encounters among the planets have
been shown to also be an effective mechanism for the capture of the
systems of irregular satellites (Nesvorny {{\textit{et al.}}},
2007). These arguments, which are different and independent of those
reported in this paper, also support a violent evolution scenarios of
the outer solar system. \\

In a forthcoming paper we will investigate the orbital dynamics of
the terrestrial planets in the context of the evolution of the giant
planets that we have outlined in this work.

\begin{acknowledgements}
This work is part of the Helmholtz Alliance's 'Planetary evolution and life', which RB and AM thank for financial support. Exchanges between Nice and Thessaloniki have been funded by a PICS programme of France's CNRS, and RB thanks the host KT for his hospitality during a recent visit. RG thanks Brasil's CNPq and FAPERJ for financial support. HFL thanks NASA's OSS and OPR programmes for support. Most of the simulations in this work were performed on the CRIMSON Beowulf cluster at OCA.
\end{acknowledgements}

\section{Bilbiography}
\v{C}uk, M. 2007, DPS 39, 60 \\
D'Angelo, G., Lubow, S. \& Bate, M., 2006 ApJ 652, 1698 \\
Fernandez, J. A. \& Ip, W.-H. 1984 Icarus 58, 109\\
Goldreich, P. \& Sari, R. 2003, ApJ 585, 1024 \\
Goldreich, P., Lithwick, Y. \& Sari, R. 2004, ApJ 614, 497 \\
Gomes, R., Morbidelli, A. \& Levison, H. 2004 Icarus 170, 492\\
Gomes, R., Levison, H., Tsiganis, K. \& Morbidelli, A. 2005 Nature 435,466\\
Hahn, J. \& Malhotra, R. 1999 AJ 117,3041\\
Henrard, J. \& Lema\^{i}tre, A. 1983 Icarus55, 482\\
Henrard, J. 1993 in Dynamics Reported New series volume 2 117\\
Kley, W. \& Dirksen, G. 2006, A\&A 447, 369 \\
Kokubo, E. \& Ida, S. 1996, Icarus 123, 180 \\
Kokubo, E. \& Ida, S. 1998, Icarus 131, 171 \\
Levison, H. \& Duncan, M. 1994 Icarus 108, 18 \\
Lubow, S. \& Ogilvie, G. 2001 ApJ 560, 997\\
Malhotra, R. 1993 Nature 365, 819 \\
Malhotra, R. 1995 AJ 110, 420\\
Masset F. \& Snellgrove, M. 2001 MNRAS 320, 55\\
Michtchenko, T. \& Malhotra, R. 2004 Icarus 168, 237\\
Michtchenko, T., Beaug\'{e}, C. \& Ferraz-Mello S. 2008 MNRAS 391, 215\\
Morbidelli, A. 2002 Modern Celestial Mechanics - Aspects of Solar System Dynamics (Taylor \& Francis, UK).
Morbidelli A. \& Crida, A. 2007 Icarus 191, 158\\
Morbidelli, A., Tsiganis, K., Crida, A., Levison, H. \& Gomes, R. 2007 AJ 134,1790 \\
Murray, C. \& Dermott, S. 1999 Solar System Dynamics (Cambridge University Press, Cambridge, UK) \\
Neishtadt, A. 1984 J. App. Math. Mech 48, 133 \\
Nelson, R. 2005 A\&A 443 1067 \\
Nesvorn\'{y}, D., Vokrouhlick\'{y}, D. \& Morbidelli, A. 2007 AJ 133, 1962 \\
Nobili, A., Milani, A. \& Carpino, M. 1989 A\&A 210, 313\\
Pierens, A. \& Nelson, R. 2008 A\&A 482, 333\\
Saslaw, W. 1985 ApJ 297, 49 \\
Stewart, G. \& Wetherill G. 1988 Icarus 74, 542\\
Tsiganis, K., Gomes, R., Morbidelli, A. \& Levison, H.  2005 Nature 435, 459\\
Thommes, E., Duncan, M. \& Levison, H. 1999 Nature 402, 635\\
Thommes, E., Nilsson, L \& Murray, N. 2007 ApJ 656, 25\\
Wisdom, J. \& Holman, M. 1991 AJ 102, 1528 
\end{document}